\documentclass{emulateapj}

\usepackage{amsmath}

\newcommand{\about}{$\sim\!\!$~}

\newcommand{\be}{\begin{displaymath}}
\newcommand{\ee}{\end{displaymath}}

\def\lsim{\hbox{\rlap{\raise 0.425ex\hbox{$<$}}\lower 0.65ex\hbox{$\sim$}}}
\def\gsim{\hbox{\rlap{\raise 0.425ex\hbox{$>$}}\lower 0.65ex\hbox{$\sim$}}}
\def\arcmin{\hbox{$^\prime$}}

\def\arcdeg{\mbox{$^\circ$}}

\shorttitle{Estimating the First-Light Time of SN 2014J in M82}
\shortauthors{Zheng et~al.}

\begin{document}

\title{Estimating the First-Light Time of the Type Ia Supernova 2014J in M82}

\author{WeiKang Zheng\altaffilmark{1,2},
Isaac Shivvers\altaffilmark{1},
Alexei V. Filippenko\altaffilmark{1},
Koichi Itagaki\altaffilmark{3},
Kelsey I. Clubb\altaffilmark{1}, 
Ori D. Fox\altaffilmark{1},
Melissa L. Graham\altaffilmark{1},
Patrick L. Kelly\altaffilmark{1}, and
Jon C. Mauerhan\altaffilmark{1},
}

\altaffiltext{1}{Department of Astronomy, University of California, Berkeley, CA 94720-3411, USA.}
\altaffiltext{2}{e-mail: zwk@astro.berkeley.edu .}
\altaffiltext{3}{Itagaki Astronomical Observatory, Teppo-cho, Yamagata 990-2492, Japan.}

\begin{abstract}
The Type~Ia supernova (SN~Ia) 2014J in M82 ($d\approx3.5$\,Mpc)
was serendipitously discovered by S.~Fossey's group on 2014~Jan.~21~UT
and has been confirmed to be the nearest known SN~Ia since at least SN~1986G.
Although SN~2014J was not discovered until $\sim7$\,days after first light,
both the Katzman Automatic Imaging Telescope at Lick Observatory
and K.~Itagaki obtained several prediscovery observations of SN~2014J.
With these data, we are able to constrain the object's time of first 
light to be Jan.~14.75~UT, only $0.82\pm0.21$\,d before our first detection.
Interestingly, we find that the light curve is well described by a 
varying power law,
much like SN~2013dy, which makes SN~2014J the second example of
a changing power law in early-time SN~Ia light curves.
A low-resolution spectrum taken on Jan.~23.388~UT, $\sim8.70$\,d after 
first light, shows that SN~2014J is a heavily reddened but otherwise
spectroscopically normal SN~Ia.
\end{abstract}


%%%%%%%%%%%%%%%%%%%%%%%%%%%%%%%%
%%  Section 1:  Introduction  %%
%%%%%%%%%%%%%%%%%%%%%%%%%%%%%%%%

\section{Introduction}\label{s:intro}

Type~Ia supernovae (SNe~Ia; see Filippenko 1997 for a review of SN classification) 
are used as standardizable candles
and therefore have many important applications, including measurements
of the changing expansion rate of the Universe
(Riess et~al. 1998; Perlmutter et~al. 1999). 
However, our understanding of the progenitor systems and explosion
mechanisms of SNe~Ia remains substantially incomplete.
It is well accepted that SNe~Ia are the product of the thermonuclear explosions of C/O white
dwarfs (Hoyle \& Fowler 1960; Colgate \& McKee 1969; see Hillebrandt \& Niemeyer 2000
for a review), but early discovery and detailed observations are essential
in order to determine the exact nature of the progenitor system and
the details of the explosion process.
Fortunately, with the modern telescopes and techniques now being used 
in searches for SNe, a number of recent SNe~Ia have been 
discovered when quite young and have been studied in detail. Examples
include SN~2009ig (Foley et~al. 2012), 
SN~2011fe (Nugent et~al. 2011; Li et~al. 2011), 
SN~2012cg (Silverman et~al. 2012a), SN~2012ht (Yamanaka et~al. 2014), and
SN~2013dy (Zheng et~al. 2013; hereafter Z13).

Before SN~2014J, the nearest SN~Ia detected in the modern age was
SN~1986G in NGC~5128 (distance $d=3.8\pm0.1$\,Mpc; Harris et~al. 2010),
or perhaps SN~1972E in NGC~5253 ($d=2.5$--8.0\,Mpc, with a mean of 
$\sim3.8$\,Mpc; e.g., Phillips et~al. 1992; Sandage \& Tammann 1975; 
Della Valle \& Melnick 1992; Branch et~al. 1994; Sandage et~al. 1994).
SN~2011fe, another SN~Ia found fewer than
3\,yr ago, occurred in the nearby galaxy M101
($d=6.4\pm0.7$\,Mpc; Shappee \& Stanek 2011).
The newly discovered SN~2014J in M82
($d=3.5\pm0.3$\,Mpc; Karachentsev \& Kashibadze 2006) is much closer
than SN~2011fe, and it also appears to be 
slightly closer than SN~1986G or SN~1972E (but the latter's large distance
uncertainty precludes an exact comparison). This means that 
SN~2014J offers researchers a unique opportunity to study
a nearby SN~Ia in detail.

In this {\em Letter} we present our prediscovery photometric observations of SN~2014J
along with an optical spectrum taken just two days after the discovery was reported.
We include an analysis of these data and 
compare the early-time light curve of SN~2014J with that of SN~2013dy, another
SN~Ia found when it was very young.
%and found they are like twins.

%%%%%%%%%%%%%%%%%%%%%%%%%%%%%%%%%%%%%%%%%%%%%%%%
%%  Section 2:  Discovery %%
%%%%%%%%%%%%%%%%%%%%%%%%%%%%%%%%%%%%%%%%%%%%%%%%

\section{Observations and Data Processing}\label{s:discovery}

SN~2014J was serendipitously discovered by astronomer Stephen J. Fossey and
a team of his students on Jan.~21.805 (UT dates are used herein)
with a 0.35\,m telescope at the University of London Observatory (Fossey et~al. 2014).
A number of prediscovery observations including detections and nondetections were
reported after the discovery information was posted (e.g., Fossey et~al. 2014;
Ma et~al. 2014; Denisenko et~al. 2014).
Among these reports, the earliest broadband detection is from the ROTSE team (Fossey et~al. 2014),
on Jan.~15.378. After this {\it Letter} was submitted, Goobar et~al. (2014) also 
reported an iPTF detection in an H$\alpha$ image taken on Jan.~15.18.

%merely 0.43\,d after first light. %haven't explained our first light
%measurement yet

SN~2014J was also observed by the 0.76\,m Katzman Automatic Imaging Telescope (KAIT) as part 
of the Lick Observatory Supernova Search (LOSS; Filippenko et~al. 2001). 
The host galaxy, M82, was monitored by KAIT with an average cadence of 2\,days (Zheng et~al. 2014)
before the reported discovery. The supernova is clearly detected in images taken on
Jan.~16.381 with no detection on Jan.~14.365 (unfiltered limiting magnitude $>18.9$).
We measure its J2000.0 coordinates to be $\alpha=09^{\mathrm{h}}55^{\mathrm{m}}42.108^{\mathrm{s}}$, 
$\delta=+69^{\circ}40\arcmin25\farcs87$, with an uncertainty of $0\farcs20$ in each coordinate.
SN~2014J is $55\farcs2$ west and $20\farcs0$ south of the somewhat ill-defined nucleus of M82.
It is unfortunate that KAIT/LOSS did not automatically discover SN~2014J when first observed by KAIT.
The paucity of suitable stars in the field, as well as the bright and complex background light
from M82, confounded our image-subtraction and object-identification pipeline.

Several images of SN~2014J were also taken by K.~Itagaki,
with a daily cadence from Jan.~13 to Jan.~17, using a 0.5\,m telescope at the
Itagaki Astronomical Observatory, 
%Teppo-cho, Yamagata, 
Japan. The object is
clearly detected in an image taken on Jan.~15.571, with no detection in an image
from Jan.~14.559 (unfiltered limiting magnitude $> 18.0$).
These data are used together with KAIT observations to perform a
joint analysis, constraining the time of first light\footnote{Throughout 
this {\it Letter} we refer to the time of first light instead of explosion
time because the SN may exhibit a ``dark phase" lasting hours to days
between the moment of explosion and the first emitted
light (e.g., Rabinak, Livne, \& Waxman 2012; Piro \& Nakar 2012, 2013).
We define the time of first light to be the time at which
the luminosity of the SN is exactly zero in our model.
As noted by Riess et~al. (1999), ``In principle, the initial
luminosity is that of a white dwarf ($M_B=10$--15 mag), but at the 
observed speed of the rise, the brightening from zero to a white-dwarf 
luminosity requires less than 1\,s."
}
of SN~2014J.
Both the KAIT and Itagaki prediscovery data were taken in unfiltered bands,
while our multi-filter observations began only after discovery;
thus, here we focus our analysis on the early unfiltered light curve.

\begin{figure}[!]
%\begin{figure}[!hbp]
\centering
%\includegraphics[width=.49\textwidth]{Fig_1a.eps}
%\includegraphics[width=.49\textwidth]{Fig_1b.eps}
%for arXiv
\includegraphics[width=.49\textwidth]{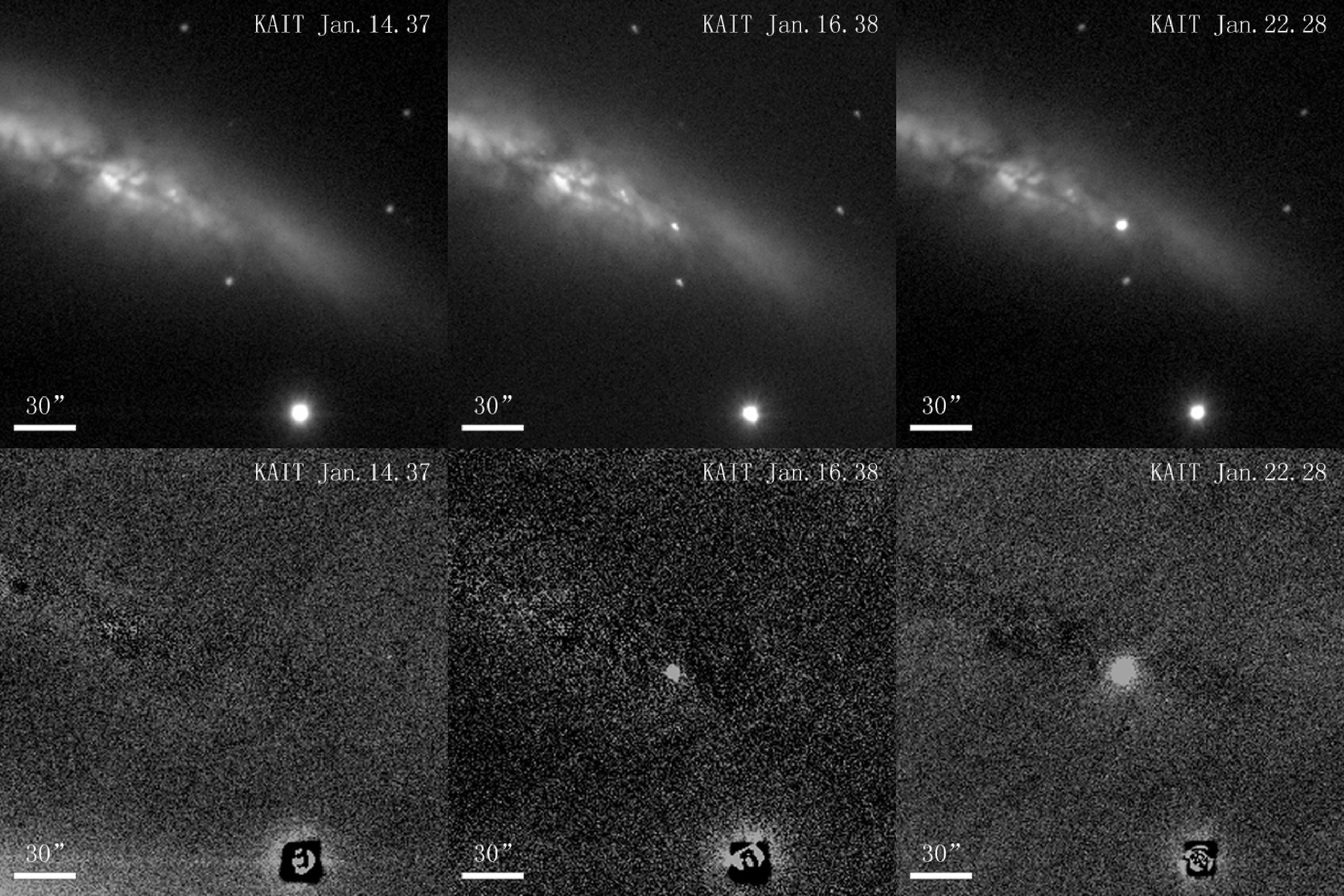}
\includegraphics[width=.49\textwidth]{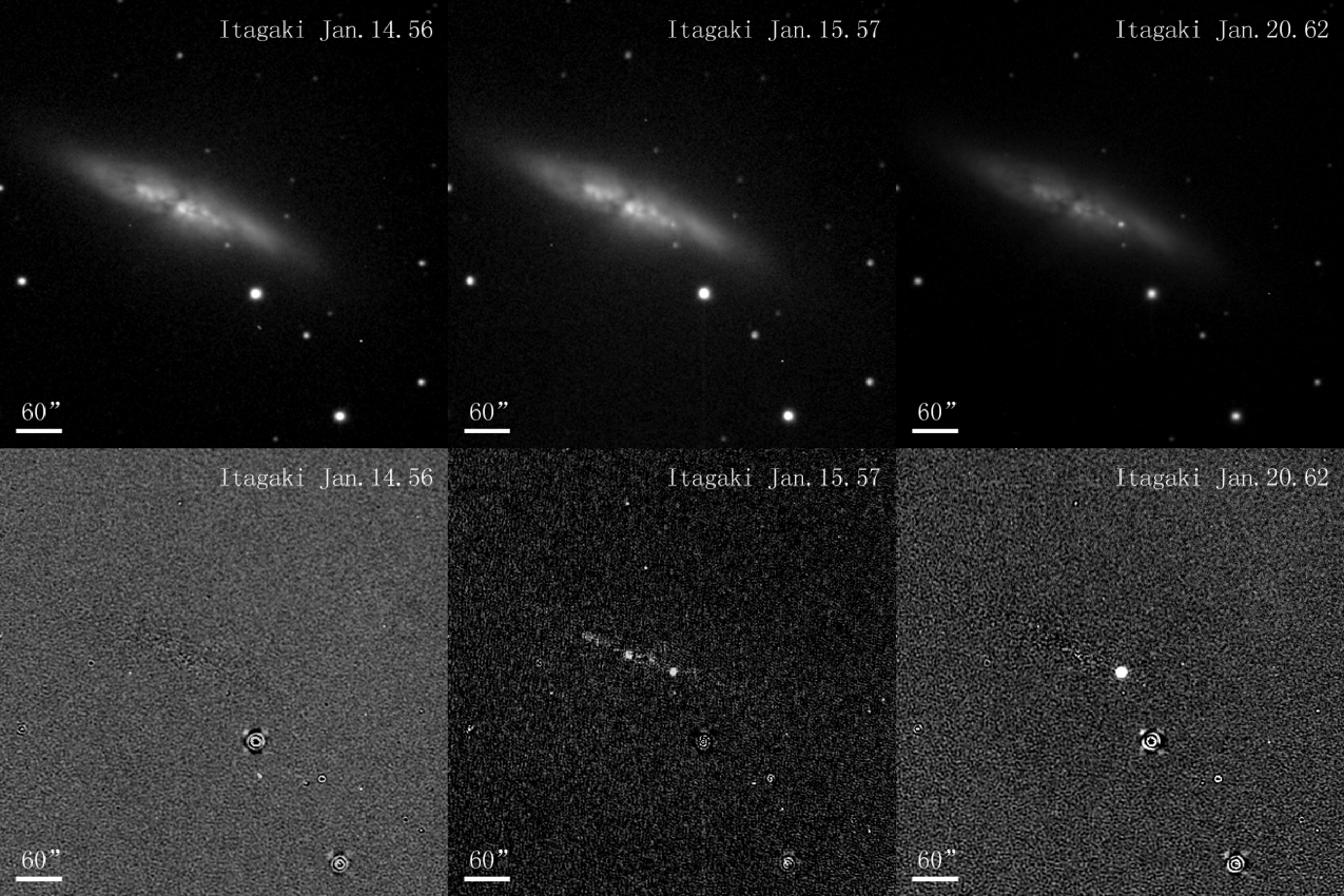}
\caption{Top row: KAIT images of SN~2014J at three different epochs,
         including the latest KAIT nondetection on Jan.~14.37.
         Second row: KAIT residual images at the same epochs
         after subtraction.
         Third and bottom rows: Same as the top two rows, but for the
         Itagaki images, including the latest nondetection on Jan.~14.56.
	SN~2014J is located at the center of each image. 
         }
\label{FC_KAIT}
\end{figure}

All KAIT images were reduced using our image-reduction pipeline (Ganeshalingam et~al. 2010),
and we similarly processed and reduced all of the Itagaki images.
In order to remove the host-galaxy contribution, we applied the same image-subtraction
procedure to both KAIT and Itagaki images using a galaxy template image taken
with each telescope before the SN explosion.
Figure \ref{FC_KAIT} shows several examples of both KAIT and Itagaki images before and
after subtraction, with SN~2014J located at the center of each image.
Point-spread-function photometry was then
performed using DAOPHOT (Stetson 1987) from the IDL Astronomy User's
Library\footnote{http://idlastro.gsfc.nasa.gov/}.
The SN instrumental magnitudes have been calibrated to two nearby stars from
the USNO B1.0 catalog: S1 with $R2 = 15.45$\,mag and J2000 coordinates
$\alpha=09^{\mathrm{h}}55^{\mathrm{m}}25.2^{\mathrm{m}}$, 
$\delta=+69^{\circ}41\arcmin21\farcs8$; S2 with $R2 = 15.09$\,mag and J2000 coordinates
$\alpha=09^{\mathrm{h}}55^{\mathrm{m}}46.1^{\mathrm{m}}$, 
$\delta=+69^{\circ}42\arcmin01\farcs8$.

For nondetections, we present an upper-limit magnitude (3$\sigma$).
Owing to the complicated background light from M82, it is inappropriate to
measure the image noise directly from the original image.
However, the host-galaxy background structure is mostly absent 
after subtraction (as shown in Fig.~\ref{FC_KAIT}).
We therefore measure the standard deviation in the residual image
at several positions around SN~2014J and use these to derive a 3$\sigma$ 
upper limit on the object's brightness and its uncertainty for each image.
This is further cross-checked by simulation: we inject a simulated
star signal (3$\sigma$ of sky noise) into the nondetection images at the SN
position, perform the same subtraction procedure, and measure the
magnitude of the simulated star in residual images using aperture photometry.
In 100 simulation trials, we are able to recover a majority ($>95$\%) of the 
measurements with a mean magnitude value within 0.1\,mag of our reported 
limiting magnitude.

\begin{figure}[!hbp]
\centering
\includegraphics[width=.49\textwidth]{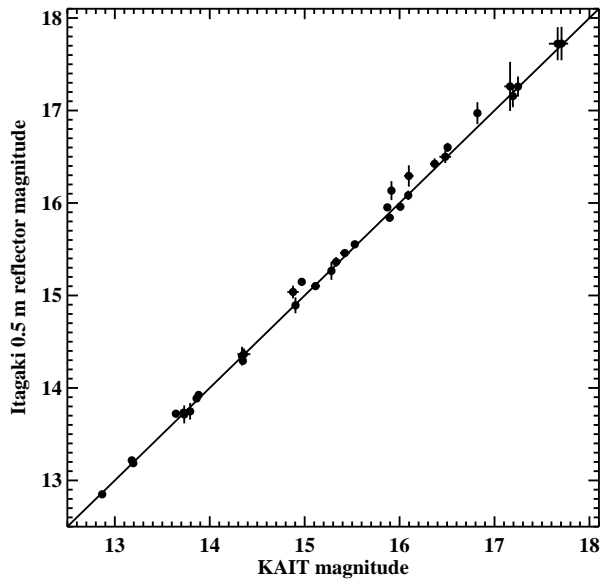}
\caption{Comparison of the unfiltered KAIT and Itagaki magnitude systems
         for stars from 4 different fields observed by
	      both telescopes at 4--15 different epochs.
         The solid line represents equivalence between the two systems.
	 We find that the Itagaki magnitudes are systematically larger (fainter)
	 by $\sim0.02$\,mag by measuring the median value of the differences.}
\label{Fig_refcompare}
\end{figure}

Our analysis includes unfiltered observations from two separate
telescopes and cameras,
so it is necessary to measure any possible offset between
KAIT and Itagaki magnitudes before beginning a
joint analysis. Since both are unfiltered, the response of each
system is dominated by the CCD quantum efficiency (QE). 
KAIT uses a MicroLine 77 camera from Finger Lakes
Instrumentation\footnote{http://www.flicamera.com/index.html}
(chip model E2V CCD77-00-BI-IMO), with a QE curve
that reaches half-peak values at \about3800 and \about8900\,\AA.
Mr.~Itagaki uses a BN-83E camera manufactured by Bitran\footnote{https://www.bitran.co.jp/}
(chip model KAF-1001E), with a QE curve that reaches
half-peak values at \about4100 and \about8900\,\AA.
These response curves, and therefore the effective passbands of the 
two systems, are very similar.
However, we can directly check by
comparing photometry of isolated field stars that are present in both datasets.
We did this for more than 30 stars in 4 different fields
(including the SN~2014J field) that have recently been observed by both KAIT and Itagaki.
We find that, overall, the unfiltered magnitudes measured from KAIT and Itagaki
have a systematic offset of only 0.02\,mag and a scatter of 0.02\,mag.
These stars exhibit a range of more than 0.9\,mag in $B-R$ color, so we
expect color-dependent difference between the two photometric systems to be small.
Figure \ref{Fig_refcompare} shows a detailed magnitude comparison between the two systems,
and Table 1 lists the raw photometry for both datasets.  Before performing our joint analysis,
we transform the Itagaki data into the KAIT magnitude system by subtracting 0.02\,mag.

%\begin{figure}[!]

\begin{deluxetable}{lcccc}
 \tabcolsep 0.4mm
 \tablewidth{0pt}
 \tablecaption{Unfiltered Photometry of SN~2014J
               \label{tab:prompt_spec_par}}
 \tablehead{ \colhead{MJD} & \colhead{UT} & \colhead{Mag} & \colhead{$1\sigma$ Error} & \colhead{Telescope} }
 \startdata
56667.4384      &    Jan. 10.4384   &   $>$18.2  & -    & KAIT     \\
56671.3653$^1$  &    Jan. 14.3653   &   $>$18.9  & -    & KAIT     \\
56673.3804      &    Jan. 16.3804   &      13.38 & 0.04 & KAIT     \\
56673.3811      &    Jan. 16.3811   &      13.37 & 0.04 & KAIT     \\
56675.3468      &    Jan. 18.3468   &      12.24 & 0.07 & KAIT     \\
56675.3471      &    Jan. 18.3471   &      12.21 & 0.04 & KAIT     \\
56675.3474      &    Jan. 18.3474   &      12.17 & 0.06 & KAIT     \\
56677.4528      &    Jan. 20.4528   &      11.27 & 0.08 & KAIT     \\
56677.4535      &    Jan. 20.4535   &      11.22 & 0.11 & KAIT     \\
56679.2793      &    Jan. 22.2793   &      10.73 & 0.04 & KAIT     \\
56679.4556      &    Jan. 22.4556   &      10.59 & 0.25 & KAIT     \\
56680.3441      &    Jan. 23.3441   &      10.39 & 0.09 & KAIT     \\
56680.3521      &    Jan. 23.3521   &      10.36 & 0.04 & KAIT     \\
56680.3852      &    Jan. 23.3852   &      10.41 & 0.07 & KAIT     \\
\\                
\hline           
\\              
56670.5914      &    Jan. 13.5914   &   $>$17.9  & -    & Itagaki  \\
56671.5588      &    Jan. 14.5588   &   $>$18.0  & -    & Itagaki  \\
56672.5705      &    Jan. 15.5705   &      14.01 & 0.03 & Itagaki  \\
56673.6414      &    Jan. 16.6414   &      13.28 & 0.06 & Itagaki  \\
56674.6124      &    Jan. 17.6124   &      12.67 & 0.13 & Itagaki  \\
56676.6179      &    Jan. 19.6179   &      11.58 & 0.06 & Itagaki  \\
56677.6205      &    Jan. 20.6205   &      11.24 & 0.05 & Itagaki  \\
\enddata
\tablenotetext{1}{Coadd of 3 individual images.}
\end{deluxetable}

An optical spectrum of SN~2014J was obtained on Jan~23.388, $\sim8.70$\,d after first light,
with the Kast double spectrograph (Miller \& Stone 1993) on the Shane 3\,m telescope at Lick Observatory. The $2''$ wide slit was aligned along the parallactic angle to
minimize the effects of atmospheric dispersion (Filippenko 1982).
The spectrum was reduced following standard techniques and was flux calibrated through observations
of appropriate spectrophotometric standard stars (e.g., Silverman et~al. 2012b).
We deredshift it into
the rest frame of M82 using $v=203$\,km\,s$^{-1}$ (de Vaucouleurs et~al. 1991).
We correct for
reddening due to Milky Way dust along the line of sight using the $R_V=3.1$ reddening law
of Cardelli, Clayton, \& Mathis (1989), adopting $E(B-V)_{\rm MW}=0.14$\,mag (Schlafly \& Finkbeiner 2011).
In addition, SN~2014J is 
substantially obscured by dust in the host galaxy, M82.  The very strong \ion{Na}{1} absorption
lines imparted on the spectrum by the interstellar medium (ISM) in M82 are saturated, and they exhibit total equivalent widths of
2.4\,\AA~and 2.7\,\AA~for D1 and D2, respectively (Cox et~al. 2014; Polshaw et~al. 2014).
These values are well beyond the range where commonly used
empirical relations for determining reddening are valid (Poznanski et~al. 2013), making any sort
of accurate host-galaxy reddening correction difficult at this time.

%%%%%%%%%%%%%%%%%%%%%%%%%%%
%%  Section 3:  Analysis & Results %%
%%%%%%%%%%%%%%%%%%%%%%%%%%%

\section{Analysis and Results}\label{s:analysis}
\subsection{Light Curves}\label{ss:lightcurves}

\begin{figure}[!hbp]
\centering
\includegraphics[width=.49\textwidth]{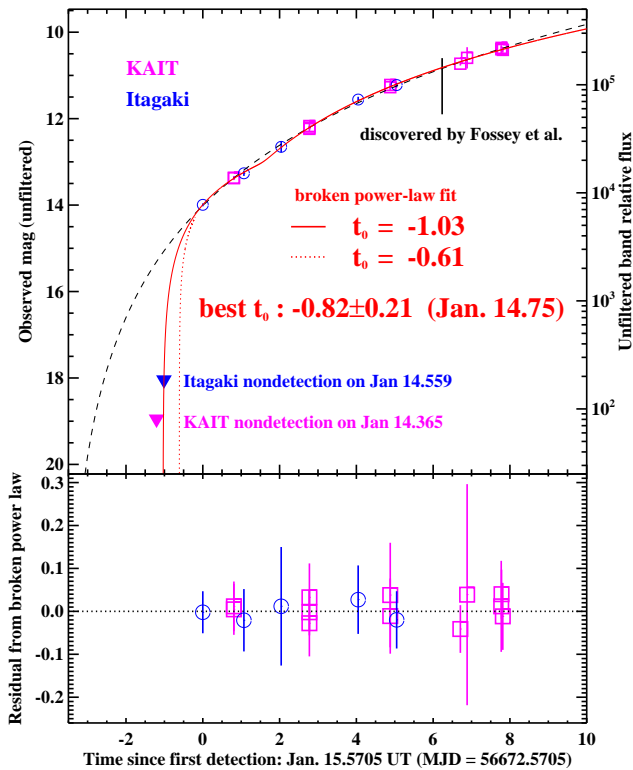}
\caption{The top panel shows a broken power-law fit to the 
  unfiltered light curve of SN~2014J
  with two different methods. Method 1 (solid red line) marks a lower bound
  on the date of first light by assuming that
  the latest nondetection (filled blue triangle)
  is a marginal ``real detection'' during fitting. Method 2 (dotted red line)
  marks an upper bound on the date of first light by 
  adopting the 3$\sigma$ limit for ${\alpha}1$ as fit
  from SN~2013dy, assuming that SN~2014J and SN~2013dy exhibited
  similar rise rates at very early times.
  See text for more details. The dashed black line is the best-fit result
  for a single power-law model, shown here for comparison.
  The bottom panel shows the residuals for Method 1.
  Our best estimate of the time of first light for SN~2014J is
  Jan.~14.75, or $0.82\pm0.21$\,d before our first detection.
  \label{figure_lc_brokenPL}}
\end{figure}

Figure \ref{figure_lc_brokenPL} shows our unfiltered light curves of SN~2014J, with
KAIT data in pink and Itagaki data in blue. Our first detection of
SN~2014J comes from an Itagaki image taken Jan.~15.571 (14.01\,mag),
followed by a KAIT detection on Jan.~16.381 (13.37\,mag). Our
latest nondetection comes from an Itagaki
image taken Jan.~14.559 ($>$18.0\,mag), with
a KAIT nondetection from three coadded images at a mean time of Jan.~14.365 ($> 18.9$\,mag).
These deep nondetections (more than 4\,mag deeper than the
first detection) within a single day of the first detection
allow us to put a very tight constraint on the time of first light.

%\begin{figure}[!]

To determine the first-light time,
one can assume that the SN luminosity scales as the
surface area of the expanding fireball, and therefore increases
quadratically with time ($L \propto t^{2}$, commonly known as the
$t^2$ model; e.g., Arnett 1982; Riess et~al. 1999). The $t^2$ model
provides a good fit for several SNe~Ia with early observations (e.g.,
SN~2011fe, Nugent et~al. 2011; SN~2012ht, Yamanaka et~al. 2014).
We applied this model to the joint dataset from both KAIT and Itagaki, 
finding a poor fit (with $\chi^2$ per degree of
freedom of 1.73); the data imply a steeper power-law index.
We therefore free the index of the power law and obtain a best-fit 
value of $2.89\pm0.27$ ($\chi^2/\nu=0.26$),
with a corresponding first-light time of Jan.~11.88, or
$3.56\pm0.58$\,d before our first detection.

Though the simple power-law model fits the detected light curve well
(dashed black line in Fig.~\ref{figure_lc_brokenPL}),
it is strongly at odds with both the KAIT and Itagaki nondetections.
The very deep nondetection limit requires the
first-light time to be much closer to our first detection point,
and a rapid increase in brightness (more than 4\,mag) within a 
short time (one day) therefore yields a different power-law index.
This is very similar to the case of SN~2013dy,
which also showed a very rapid rise (the brightness of SN 2013dy increased more
than 2\,mag within the first 0.5\,d). 
Hence, in a manner similar to our work on SN~2013dy (Z13), as well
as to models widely used for observed gamma-ray burst afterglows
(e.g., Zheng et~al. 2012), we adopt 
a broken power-law model with a variable index 
%to fit the rapid rise of the early-time light curve of 
for SN~2014J:
\begin{equation}
f = \left(\frac{t-t_0}{t_b}\right)^{{\alpha}1} \Big{[} 1 +
\left(\frac{t-t_0}{t_b}\right)^{s({\alpha}1-{\alpha}2)}\Big{]}^{-1/s},
\end{equation}
where $f$ is the flux, $t_0$ is the first-light time, $t_b$ is the break 
time, ${\alpha}1$ and ${\alpha}2$ are the power-law indices before 
and after the break (respectively), and $s$ is a smoothing parameter.

In Z13, we were able to constrain the value of ${\alpha}1$ with a 
very early detection of SN~2013dy; sadly, we do not have a similarly early
detection of SN~2014J.  However, we use two methods to obtain quite narrow 
limits on the first-light time with the existing data.
In Method 1, we treat the latest nondetection as a marginal ``real detection,"
yielding an effective upper limit on the value of ${\alpha}1$.
Assuming our broken power-law model applies back to the time of first light,
it also provides a lower limit on this time.
Based on this assumption, and setting the time of our first detection to $t=0$, our best fit gives 
$t_0=-1.03$\,d, $t_b=2.62\pm0.22$\,d, $s=-36.9\pm12.3$, ${\alpha}1=0.97\pm0.11$, and
${\alpha}2=1.98\pm0.07$ ($\chi^2/\nu=0.16$).
This implies that first light occurred on Jan.~14.54, and
the fit is shown by the solid red line in Figure \ref{figure_lc_brokenPL}.
In Method 2, we assume that the very early rise of SN~2014J was similar to that
of SN~2013dy.  The best-fit ${\alpha}1$ value found for SN~2013dy was $0.88\pm0.07$ (Z13).
To place a conservative upper limit on the first-light time of SN~2014J, we adopt 
${\alpha}1=0.67$, the ${\alpha}1$ value (3$\sigma$) from SN~2013dy.
We find a best-fit model with
$t_0=-0.61$\,d, $t_b=2.20\pm0.12$\,d, $s=-202.8\pm67.6$, ${\alpha}1=0.67$ (fixed during fitting), and
${\alpha}2=1.82\pm0.07$ ($\chi^2/\nu=0.17$).
This implies that first light occurred on
Jan.~14.96, and the fit is shown by the dotted red line in Figure \ref{figure_lc_brokenPL}.

As shown above, the best fits for the $t^{2.89}$ model and the two broken
power-law models are much better
than the $t^2$ model. The $\chi^2/\nu$ value for the $t^{2.89}$ model and the
two broken power-law models are comparable, but a simple $\chi^2/\nu$ analysis does not take into
account the strong constraints from our nondetections, which rule out a simple power-law model.
We therefore favor the broken power-law model.

Our two methods constrain
the time of first light to be somewhere
between $1.03$\,d and $0.66$\,d before our first detection.
We adopt, as our best estimate, the mean of these two values:
Jan.~14.75, or $0.82\pm0.21$\,d before our first detection.
Both the iPTF-H$\alpha$ and ROTSE detections are after, 
and thus consistent with, our derived first-light time 
(by 0.43\,d and 0.63\,d, respectively).
This makes SN~2014J one of the earliest detected SNe~Ia, along with
SN~2013dy (0.10\,d after first light; Z13),
SN~2011fe (0.46\,d; Nugent et~al. 2011), and SN 2009ig 
(0.71\,d; Foley et~al. 2012).

The extremely rapid rise in the first day of SN~2014J's light curve
reinforces several of the conclusions obtained by Z13 when studying SN~2013dy,
showing that the $t^2$ model is not sufficient for every SN~Ia.
(1) Within the first day after
first light, some SNe~Ia exhibit a very rapid increase in brightness, in the case
of SN~2014J becoming more than 4\,mag brighter in a single day. 
Actually observing this very rapid rise is a challenge given the very short timespan
over which it occurs.
(2) Some SN~Ia light curves are best described by power laws with an exponent 
not equal to 2 (see also Piro \& Nakar 2012).
(3) The best-fit power-law exponent likely varies with time.
With SN~2014J, we add to the mounting evidence that the $t^2$ model has worked for previous SNe~Ia 
only because observations constraining the shape of the light curve at very early times
were rare.  The early-time light curves of
SN~2013dy and SN~2014J demonstrate that a varying power-law
exponent may be a common phenomenon in the light curves of very young SNe~Ia,
and that SNe~Ia are likely more complex
than the simple fireball model assumes.

The physical explanation for the varying exponent of SNe~Ia is still unclear.
Perhaps the very early fireball exhibits significant changes in either the photospheric
temperature or the velocity during expansion, or the fireball input energy may change
owing to the geometric structure of the Ni$^{56}$ distribution in outer layers.
Another possibility is that the very early light curve may include some
contribution from the shock-heated cooling emission after shock 
breakout, although this phenomenon is predicted to exhibit a power-law index of 1.5 
($f \propto t^{1.5}$; see Eq.~3 of Piro \& Nakar 2013), not in good agreement
with the values observed in SN~2013dy and SN~2014J. 

\subsection{Optical Spectra}\label{ss:spectra}

Our Lick spectrum taken on Jan.~23.39 shows that SN~2014J is a spectroscopically
normal SN~Ia showing some high-velocity features and strong dust reddening,
as noted by others (e.g., Cao et~al. 2014).  
The analysis above indicates that this spectrum was taken $\sim 8.70$\,d after 
first light.  Note the prominent narrow absorption features produced by the 
host-galaxy ISM:
two \ion{Ca}{2} lines near 3950\,\AA\ and the \ion{Na}{1} line near 5900\,\AA.
We classify the spectrum with the SuperNova IDentification code
using an enhanced set of spectral templates ({\tt SNID}; Blondin \& Tonry 2007; 
Silverman et~al. 2012b),
which indicates a 100\% match with premaximum SN~Ia spectra ($\sim90$\% match 
to the SN~Ia-norm subtype and $\sim10$\% match to the SN~Ia-99a subtype).

\begin{figure*}[!ht]
\centering
\includegraphics[width=0.9\textwidth]{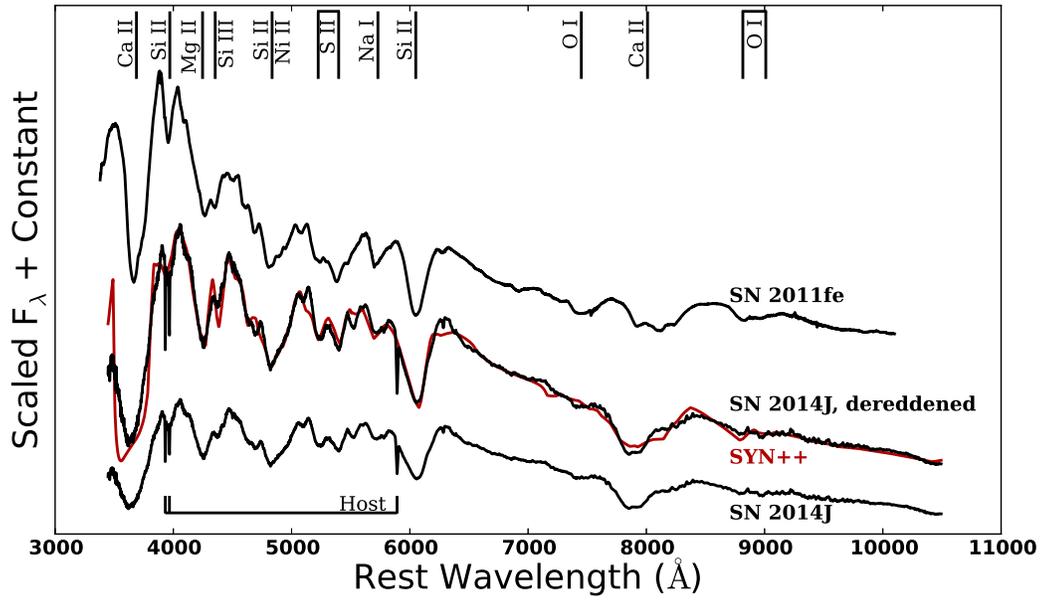}
\caption{Spectrum of SN~2014J taken $\sim8.70$\,d after first light.
         The bottom spectrum shows SN~2014J after applying a reddening correction
         only for Milky Way dust. 
         The middle spectrum displays SN~2014J
         after a reddening correction assuming a total $E(B-V)=0.8$\,mag, with
          our best {\tt SYN++} fit given in red.  The top spectrum shows SN~2011fe
         at $\sim7.5$\,d after first light, for comparison (Parrent et~al.~2012).  
         Major spectral features are labeled at the top.
         %All spectra are shown in the rest frame.
\label{figure_spectra} }
\end{figure*}

As mentioned in \S~\ref{s:discovery}, it is difficult to determine the exact 
amount of reddening produced by the host galaxy, M82.  However, Polshaw et~al. (2014)
suggest that the early-time light curve of SN~2014J indicates a total $E(B-V)\lesssim0.8$\,mag.
Thus, for visual comparison only, we deredden our spectrum of SN~2014J to account for
a dust contribution from M82 of $E(B-V)_{\rm M82}=0.66$\,mag, assuming the same 
dust-law parameterization as described in \S \ref{s:discovery} (in addition to the 
already-applied correction for Milky Way extinction).  
We then use {\tt SYN++}/{\tt SYNAPPS} (Thomas et~al. 2011)
to fit a simple parameterized SN~Ia model to the dereddened spectrum.  We obtain a 
good fit using only ions commonly found in SN~Ia spectra:
\ion{O}{1}, \ion{Na}{1}, \ion{Mg}{2}, \ion{Si}{2}, \ion{Si}{3}, \ion{S}{2}, \ion{Ca}{2},
\ion{Fe}{2}, \ion{Fe}{3}, and \ion{Ni}{2}.  We see little evidence for the presence of unburned \ion{C}{2}
at this time.
Note that we needed to include nonzero warping parameters
in our fit to {\tt SYN++},  indicating that the applied reddening correction is not exact.

Figure \ref{figure_spectra} displays our spectrum of SN~2014J after
correction for only Milky Way extinction and for an assumed total
$E(B-V)=0.8$\,mag, a spectrum of the SN~Ia-norm SN~2011fe at very
nearly the same phase $\sim7.5$\,d after first light (Parrent
et~al.~2012; Yaron \& Gal-Yam 2012), and our best-fit {\tt SYN++}
model.  Comparing the spectra of these two SNe, and our {\tt SYN++}
fit and those of SN~2011fe (Parrent et~al.~2012), two major
differences appear: SN~2014J exhibits significantly less \ion{O}{2}
and \ion{C}{2} absorption than SN~2011fe at the same phase, indicating
that SN~2014J has very little unburned material in its atmosphere at
8.7\,d, and SN~2014J shows strong evidence for high-velocity
components in several species including \ion{Ca}{2} and \ion{Si}{2}.
%An impressive observational effort is currently underway by groups
%throughout the world to collect spectra of SN~2014J; 
For now, we postpone any further spectral analysis.

%%%%%%%%%%%%%%%%%%%%%%%%%%%
%%  Section 4:  Conclusion  %%
%%%%%%%%%%%%%%%%%%%%%%%%%%%

\section{Conclusions}\label{s:conclusions}

In this {\it Letter} we present optical photometry and spectroscopy of
the normal Type Ia SN~2014J.  Despite the fact that it was found
lamentably late for such a nearby SN, we show that existing
prediscovery observations of SN~2014J constrain the first-light time
to be between Jan.~14.54 and 14.96, and we present Jan.~14.75~UT as
our best estimate.  In addition, the early-time light curve of
SN~2014J does not match the canonical $t^2$ fireball model, instead
exhibiting a variable power-law index similar to that derived for
SN~2013dy.  We look forward to further studies of this exciting
object, and we hope that it will help us to better understand the
underlying nature of SNe~Ia.
\\

\begin{acknowledgments}

A.V.F.'s group (and KAIT) at UC Berkeley have received financial
assistance from the TABASGO Foundation, the Sylvia \& Jim Katzman
Foundation, the Christopher R. Redlich Fund, Gary and Cynthia Bengier,
the Richard and Rhoda Goldman Fund, Weldon and Ruth Wood, and NSF
grant AST-1211916.
We thank the staffs at Lick Observatory and Itagaki Observatory 
where data were obtained.
%We thank the anonymous referee for the useful suggestions that
%improved the paper.

\end{acknowledgments}


\begin{thebibliography}{50}
\expandafter\ifx\csname natexlab\endcsname\relax\def\natexlab#1{#1}\fi

\bibitem[]{} Arnett, W. D. 1982, ApJ, 253, 785

%\bibitem[]{} Blanchard, P., Zheng, W., Cenko, S. B., Li, W., et al. 2013, CBET, 3422
%
\bibitem[{{Blondin} \& {Tonry}(2007)}]{Blondin07}
{Blondin}, S., \& {Tonry}, J.~L. 2007, \apj, 666, 1024

\bibitem[]{} Branch, D., et~al. 1994, ApJ, 421, 87

%\bibitem[]{} Brown, T. M., et al. 2013, arXiv:1305.2437
%

\bibitem[]{} Cao, Y., et~al. 2014, The Astronomer's Telegram, 5786, 1

\bibitem[]{} Cardelli, J. A., Clayton, G. C., \& Mathis, J. S. 1989, ApJ, 345, 245

%\bibitem[]{} Casper, C., Zheng, W., Li, W., Filippenko, A. V., \& Cenko, S. B., 2013, CBET, 3588
%
%\bibitem[]{} Childress, M., et al. 2013, ApJ, 770, 29
%
\bibitem[]{} Colgate, S. A., \& McKee, C. 1969, ApJ, 157, 623

\bibitem[]{} Cox, N. L. J, Davis, P., Patat, F., \& Van Winckel, H.  2014,
The Astronomer's Telegram, 5797, 1

\bibitem[]{} Della Valle, M., \& Melnick, J. 1992, A\&A, 257, L1

\bibitem[]{} Denisenko, D., et~al. 2014, The Astronomer's Telegram, 5795, 1

\bibitem[]{} de Vaucouleurs, G., et~al. 1991, Third Reference Catalogue of Bright Galaxies

%\bibitem[{{Faber} {et~al.}(2003){Faber}, {Phillips}, {Kibrick}, {Alcott},
%  {Allen}, {Burrous}, {Cantrall}, {Clarke}, {Coil}, {Cowley}, {Davis}, {Deich},
%  {Dietsch}, {Gilmore}, {Harper}, {Hilyard}, {Lewis}, {McVeigh}, {Newman},
%  {Osborne}, {Schiavon}, {Stover}, {Tucker}, {Wallace}, {Wei}, {Wirth}, \&
%  {Wright}}]{faberdeimos03}
%{Faber}, S.~M., {et~al.} 2003, SPIE, 4841, 1657
%

\bibitem[]{} Filippenko, A. V. 1982, PASP, 94, 715

\bibitem[]{} Filippenko, A. V. 1997, ARAA, 35, 309

\bibitem[{{Filippenko} {et~al.}(2001){Filippenko}, {Li}, {Treffers}, \&
  {Modjaz}}]{Filippenko01}
 {Filippenko}, A.~V., {Li}, W.~D., {Treffers}, R.~R., \& {Modjaz}, M. 2001, in
 {Small-Telescope Astronomy on Global Scales.}, ed. B.~{Paczy\'{n}ski}, W.~P.
 {Chen}, \& C.~{Lemme} (San Francisco: ASP), 121

\bibitem[{{Foley} {et~al.}(2012)}]{Foley12} {Foley}, R.~J., {et~al.} 2012, \apj, 744, 38

\bibitem[]{} Fossey, S. J., et~al. 2014, CBET, 3792

\bibitem[{{Ganeshalingam} {et~al.}(2010)}]{Ganeshalingam10:phot_paper}
{Ganeshalingam}, M., {et~al.} 2010, \apjs, 190, 418

%\bibitem[]{} Garavini, G., et al. 2005, AJ, 130, 2278
%
\bibitem[]{} Goobar, A., et~al. 2014, arXiv:1402.0849

\bibitem[]{} Harris, G. L. H., Rejkuba, M., \& Harris, W. E. 2010, PASA, 27, 457

%\bibitem[]{} Hicken, M., et al. 2007, ApJ, 669, L17
%
%\bibitem[{{Hill} {et~al.}(1998)}]{Hill98}
%{Hill}, G.~J., {et~al.} 1998, SPIE, 3355, 375
%
\bibitem[]{} Hillebrandt, W., \& Niemeyer, J. C. 2000, ARA\&A, 38, 191

%\bibitem[]{} Howell, D. A., et al. 2006, Nature, 443, 308 
%
\bibitem[]{} Hoyle, F., \& Fowler, W. A. 1960, ApJ, 132, 565

\bibitem[]{} Karachentsev, I., \& Kashibadze, O. G. 2006, Astrophys., 49, 3

%\bibitem[{{Kasen}(2010)}]{Kasen10} {Kasen}, D. 2010, \apj, 708, 1025
%
%\bibitem[]{} Kelly, P. L., et al. 2010, ApJ, 715, 743
%
%\bibitem[]{} Kumar, S., Fuller, K., Zheng, W., et al. 2013, CBET, 3561
%
%\bibitem[{{Li} {et~al.}(2001){Li}, {Filippenko}, {Treffers}, {Riess}, {Hu}, \&
%  {Qiu}}]{Li01:pec}
%{Li}, W., {Filippenko}, A.~V., {Treffers}, R.~R., {Riess}, A.~G., {Hu}, J., \&
%  {Qiu}, Y. 2001, \apj, 546, 734
%
\bibitem[]{} Li, W., et~al. 2011, Nature, 480, 348

\bibitem[]{} Ma, B., et~al. 2014, The Astronomer's Telegram, 5797, 1

%\bibitem[]{} Maund, J. R., et al. 2013, MNRAS, 433, L20
%
\bibitem[{{Miller} \& {Stone}(1993)}]{Miller93}
{Miller}, J.~S., \& {Stone}, R.~P.~S. 1993, {Lick Obs. Tech. Rep. 66} (Santa
  Cruz: Lick Obs.)

\bibitem[{{Nugent} {et~al.}(2011)}]{Nugent11} {Nugent}, P.~E., {et~al.} 2011, \nat, 480, 344

%\bibitem[]{} Parrent, J. T., et~al. 2011, ApJ, 732, 30

\bibitem[{{Parrent} {et~al.}(2012)}]{Parrent12}
{Parrent}, J.~T., {et~al.} 2012, \apjl, 752, 26

%\bibitem[]{} Patat, F., Benetti, S., Cappellaro, E., et al. 1996, MNRAS, 278, 111
%
%\bibitem[]{} Paturel, G., et al. 2000, A\&AS, 144, 475
%
\bibitem[{{Perlmutter} {et~al.}(1999)}]{Perlmutter99}
{Perlmutter}, S., {et~al.} 1999, \apj, 517, 565

\bibitem[]{} Phillips, M. M., et~al. 1992, BAAS, 24, 749

\bibitem[]{} Piro, A., \& Nakar, E. 2012, arXiv:1211.6438

\bibitem[]{} Piro, A., \& Nakar, E. 2013, ApJ, 769, 67

\bibitem[]{} Polshaw, J., et~al.  2014, The Astronomer's Telegram, 5816, 1

\bibitem[]{} Poznanski, D., Prochaska, J. X., \& Bloom, J. S. 2012, \mnras, 426, 1465

%\bibitem[{{Poznanski} {et~al.}(2011){Poznanski}, {Ganeshalingam}, {Silverman},
%  \& {Filippenko}}]{Poznanski11}
%{Poznanski}, D., {Ganeshalingam}, M., {Silverman}, J.~M., \& {Filippenko},
%  A.~V. 2011, \mnras, 415, L81
%
\bibitem[]{} Rabinak, I., Livne, E., \& Waxman, E. 2012, ApJ, 757

\bibitem[{{Riess} {et~al.}(1999){Riess}, {Filippenko}, {Li}, \&
  {Schmidt}}]{Riess99:risetime}
{Riess}, A.~G., {Filippenko}, A.~V., {Li}, W., \& {Schmidt}, B.~P. 1999, \aj,
  118, 2675

\bibitem[{{Riess} {et~al.}(1998)}]{Riess98:lambda}
{Riess}, A.~G., {et~al.} 1998, \aj, 116, 1009

\bibitem[]{} Sandage, A., \& Tammann, G. A. 1975, ApJ, 196, 313

\bibitem[]{} Sandage, A., et~al. 1994, ApJ, 423, L13

\bibitem[]{} Schlafly, E. F., \& Finkbeiner, D. P. 2011, ApJ, 737, 103

%\bibitem[]{} Scalzo, R. A., et al. 2010, ApJ, 713, 1073
%
%\bibitem[{{Schlegel} {et~al.}(1998){Schlegel}, {Finkbeiner}, \& {Davis}}]{Schlegel98}
%{Schlegel}, D.~J., {Finkbeiner}, D.~P., \& {Davis}, M. 1998, \apj, 500, 525
 
\bibitem[]{} Shappee, B. J., \& Stanek, K. Z. 2011, ApJ, 733, 124

%\bibitem[]{} Silverman, J. M., et al. 2011, MNRAS, 410, 585
%
\bibitem[{{Silverman} {et~al.}(2012{\natexlab{a}})}]{Silverman12:2012cg}
{Silverman}, J.~M., {et~al.} 2012a, \apj, 756, L7

\bibitem[{{Silverman} {et~al.}(2012{\natexlab{b}})}]{Silverman12:BSNIPI}
{Silverman}, J.~M., {et~al.} 2012b, \mnras, 425, 1917

%\bibitem[{{Silverman} {et~al.}(2012{\natexlab{c}})}]{}
%{Silverman}, J.~M., {et~al.} 2012c, \mnras, 425, 1819
%
\bibitem[]{} Stetson, P. B. 1987, PASP, 99, 191

%\bibitem[]{} Taubenberger, S., et al. 2011, MNRAS, 412, 2735

\bibitem[{{Thomas} {et~al.}(2011){Thomas}, {Nugent}, \&
 {Meza}}]{Thomas11:synapps}
{Thomas}, R.~C., {Nugent}, P.~E., \& {Meza}, J.~C. 2011, \pasp, 123, 237

%\bibitem[]{} Tully, R., et al., 2009, AJ, 138, 323
%
%\bibitem[]{} Wang, X., et al. 2009, ApJ, 697, 380
%
%\bibitem[]{} Wang, X., et al. 2013, Science, 340, 170
%
%\bibitem[]{} Yamanaka, M., et al. 2009, ApJ, 707, L118
%
\bibitem[]{} Yamanaka, M., et~al. 2014, arXiv:1401.5160

\bibitem[]{} Yaron, O., \& Gal-Yam, A. 2012, \pasp, 124, 668

\bibitem[]{} Zheng, W., \& Filippenko, A. V. 2014, The Astronomer's Telegram, 5828, 1

\bibitem[]{} Zheng, W., et~al. 2012, ApJ, 751, 90

\bibitem[]{} Zheng, W., et~al. 2013, ApJ, 778, L15 (Z13)

\end{thebibliography}
\end{document}